# A comparison between central- and self-dispatch storage management principles in island systems

Georgios N. Psarros[1], Pantelis A. Dratsas, Stavros A. Papathanassiou

*School of Electrical and Computer Engineering, National Technical University of Athens (NTUA), Athens, Greece*

**ABSTRACT**

This paper presents a comparative evaluation of central and self-dispatch management concepts for battery energy storage (BES) facilities in island power systems with a high renewable energy source (RES) penetration. BES facilities deployed to support the integration of additional wind capacity can be either centrally dispatched by the island System Operator or they can be self-dispatched within a Virtual Power Plant entity comprising renewables and storage, called a Hybrid Power Station (HPS). To explore the anticipated benefits of each BES management paradigm, annual simulations are performed for an example island system, employing a three-layer mixed integer linear programming (MILP) method to simulate the unit commitment and economic dispatch processes. The levelized cost of energy (LCOE) of combined BES and renewables investments is calculated and the achieved RES penetration levels and island system generation cost are evaluated, allowing the identification of Pareto optimal storage configurations, leading to lowest LCOE for a given RES penetration target. Overall, the centrally dispatched BES systems prove to be substantially more cost-effective, compared to the self-dispatched alternative, for achieving similar RES penetration levels.

*Keywords:* battery energy storage; centrally dispatched storage; self-dispatched storage; hybrid power stations; Pareto front; renewable energy

## 1. Introduction

Energy storage for grid-scale applications has been investigated extensively in the literature (and has been proven in real-world applications worldwide ([1–6]). Storage stations provide a wide spectrum of services to power systems, such as energy arbitrage ([7]), fast response reserves ([8]), black start capabilities ([9]) and several others essential to achieve smooth integration of renewable energy sources (RES) at high penetration levels, [10].

Autonomous power systems, such as non-interconnected islands or isolated microgrids, present particularities compared to large continental grids, being more vulnerable to the stochastic variations of RES production due to their inherent low inertia and lack of interconnections, [11]. The integration of intermittent RES, such as wind and solar, in the energy mix of island systems rarely exceeds 20-25% of their annual load demand, [12]. Limitations preventing further increase of RES penetration levels can be security-related (e.g. response to severe disturbances, such as the loss of large online conventional units or RES production) or imposed by the operational characteristics and flexibility

---

[1] Corresponding author

*e-mail addresses:* gpsarros@mail.ntua.gr (G. N. Psarros), pantelisdratsas@mail.ntua.gr (P. A. Dratsas), st@power.ece.ntua.gr (S. A. Papathanassiou)

limitations of thermal units (such as the minimum production level and other dispatch constraints, [13,14]). The introduction of fast-response battery energy storage (BES) systems can relax such constraints and provide the flexibility required to accommodate increased amounts of variable renewable generation in island systems ([15,16]).

Although several deployment and management concepts are available to integrate storage in isolated systems, two main paradigms emerge in the literature as regards grid-scale storage in islands, denoted hereinafter as the *central-dispatch* and the *self-dispatch* concepts. In the former, storage facilities constitute system assets, rather than independent market participants, without being associated with specific RES stations; thus, they are managed directly by the island System Operator (SO) over the entire generation scheduling process, from day-ahead scheduling (DAS) to real-time operation, [15–18]. In the *self-dispatch* concept, on the other hand, storage participates in the local market aggregated with renewable generation in Virtual Power Plant (VPP) entities, called Hybrid Power Stations (HPS) ([19–22]). In this hybrid concept, storage facilities are managed together with RES generation in portfolio mode, i.e. as a fully dispatchable aggregate entity. Intermittency and variability of RES production is intrinsically compensated via self-dispatch of the VPP components in real-time operation by the HPS-operator (HPS-o), subject to dispatch orders issued by the island SO for the entire HPS ([23]). Notably, besides their participation in the island energy market as fully dispatchable generation entities, HPS also contribute to resource adequacy ([19]) and constitute an attractive proposal for island power systems, with real-world applications in operation, [24,25].

Benefits from the introduction of centrally managed BES stations in island systems have already been examined in the literature. In [17], the introduction of centrally managed BES in two Spanish islands of different sizes reduced significantly the production cost through energy arbitrage on the daily load curve. Similarly, in [16] the impact from the introduction of BES stations on the economic operation of a medium sized island was also found to be positive, however, in this case, due to the provision of fast reserves, rather than through arbitrage. In [26,27] the island system of Cyprus was investigated in the presence of centrally managed BES of different technologies (Li-ion, Pb-acid, Zn-air, Na-S, etc.), leading to the conclusion that storage may significantly contribute to system operating reserves requirements and improve system economic operation under high RES penetration levels. In [28–30], the main objective of storage was the enhancement of RES penetration, without addressing storage investment feasibility. The main finding was that annual renewable penetration levels could reach up to 100% in the presence of suitably sized storage. Topics related to frequency regulation, including provision of inertial and fast response reserves for low- or no-inertia microgrids and isolated systems are the subject of [31–33].

Hybrid RES-storage concepts have received a lot of attention in the literature as instruments to decarbonize island systems. In [34], a HPS design is presented to achieve high RES penetration rates, over 75%, in the island of Ikaria. In [35,36], a similar concept, involving pumped hydro storage and wind, is investigated for the island of El Hierro to achieve annual RES penetration levels around 80%. The same island is the study case of [37], where a novel generation scheduling model has been introduced to cope with the intra-hourly variability of the system's residual load. Hybrid storage-renewable stations have also been evaluated in [38] as a means to achieve full decarbonization in an economically feasible manner for the Ometepe island. In [22], several HPS configurations are examined for a small island, leading to RES penetrations around 45%. HPS market offer strategies [39], component sizing and location [40], real-time self-scheduling [41], as well as issues related with system dynamic response [23,42], have also been extensively analyzed in the literature.

While aspects related to system performance improvement and RES integration have received significant attention so far, the quantification of system-level economic benefits and the feasibility of storage investments still lack a similar level of attention and thorough treatment. Specifically, a direct comparison of the central- and self-dispatch storage management concepts has not been attempted so far, to reveal which concept is more effective to enhance RES integration in a cost-optimal manner. Such a comparison is undertaken in this paper, using a medium-sized island system as a study case. To establish an equal comparison basis, combined investments in li-ion BES systems and wind farms (WFs) are examined, managed under either concept, aiming to achieve similar RES penetration levels. The economic efficiency of each solution is measured using the levelized cost of energy (LCOE), which reflects the required remuneration of the combined wind-storage investments per MWh of produced energy to achieve feasibility, as well as the variable generation cost of the entire system.

A three-layer generation scheduling problem is implemented, built upon the mixed integer linear programming (MILP) approach, to simulate the operation of the system in the presence of storage and renewables. The model can accommodate centrally dispatched storage systems and self-dispatched HPS plants, has a 24-h look-ahead horizon and is successively executed over the course of a year. Annual simulations of system operation are conducted for numerous BES-RES scenarios, leading to different wind energy penetrations and generation costs, allowing to seek RES penetration-LCOE Pareto optimality.

The methodology and outcomes of the paper are applicable to the majority of island systems with sizes ranging from a few MW up to several hundred MW, where the generation management principles adopted here remain valid. Differentiated analysis approaches may be needed at the extremes of very small systems experiencing very high RES penetration rates, where the thermal units may be off-line most of the time, and in very large systems, with organized and competitive electricity markets in place.

The remaining of this paper is organized as follows. In Section 2 the *central-dispatch* and *self-dispatch* storage management principles are discussed. Section 3 describes the case-study island and the assumptions of the analysis. In Section 4, the employed modelling and simulation approach is presented. Results are provided and discussed in Section 0. Conclusions are summarized in Section 6.

## 2. Market Structure & Storage Management Concepts

### 2.1. Market Structure in non-interconnected island systems

The generating portfolio of an island system typically consists of oil-fired conventional units, intermittent renewables and possibly storage, either standalone or aggregated in HPSs. Thermal plants are typically owned by a single entity, [43], while independent power producers may exist operating RES generation facilities remunerated based on feed-in-tariffs (FiT), [12].

As island systems lack the necessary conditions to justify establishment of fully competitive energy markets, generation scheduling is usually performed on a cost optimal basis, [12,43]. This task is performed by the independent SO, responsible for the cost-optimal unit commitment and economic dispatch (UC-ED) of all generating assets, including thermal, renewables and storage (standalone or HPS), as well as for market clearing.

Objective of the UC-ED process is the simultaneous minimization of the variable operating cost of the system and the maximization of RES penetration, subject to constraints ensuring secure operation of the system in normal conditions and in case of disturbances. Conventional units participate in UC-ED based on their actual variable generation cost, without submitting priced offers, while renewables enjoy dispatch priority against thermal generation. Security constraints often lead to curtailments of renewable production, either during the dispatch stage or in real-time operation, as described in detail in [14,44].

Due to the absence of competition and priced generation offers by thermal units, market-clearing in islands is not based on a system marginal price. Rather, thermal units are remunerated on a cost-of-service basis, allowing recovery of their actual variable generation cost and their fixed capital and operating expenses. Renewable units and HPSs are remunerated for their energy production based on FiTs, which are supposed to reflect the LCOE of each technology. Ancillary services markets do not exist, while dispatchable RES stations, such as the HPSs, receive dispatch orders by the SO and are subject to imbalance penalties in case of non-compliance.

The remuneration model of centralized storage investments in island systems remains a rather obscure topic in the relevant literature. While storages operating in organized markets may receive a variety of income streams, from their participation in forward, day-ahead, intraday, balancing markets and auction-based provision of a variety of grid services, in the case of medium and small sized islands such markets do not exist, leaving a gap in terms of potential revenue streams that enable investment feasibility. In this paper, our assumption is that investments in storage are realized to support the integration of new renewable generation and therefore storage and renewables can be treated as a combined investment characterized by an aggregate LCOE. This provides an objective basis to quantify generation cost and compare alternative configurations and management concepts.

## 2.2. Centrally dispatched storage management

Under the centralized storage deployment paradigm, the BES facilities of the system are dispatched directly by the SO, along with the other energy production units of the island, as shown in Fig. 1(a). The SO issues dispatch orders for production and reserves to all dispatchable generation and storage, as well as setpoints to RES stations to curtail their production (maximum output dispatch orders), when necessary for security considerations.

In this concept, BES are treated as system-level flexibility assets, used by the SO to optimize operation and minimize generation cost during UC-ED and real-time operation, rather than as independent assets managed by a third-party. Nevertheless, ownership of storages can belong to an independent entity, as long as they are remunerated for their services to the system.

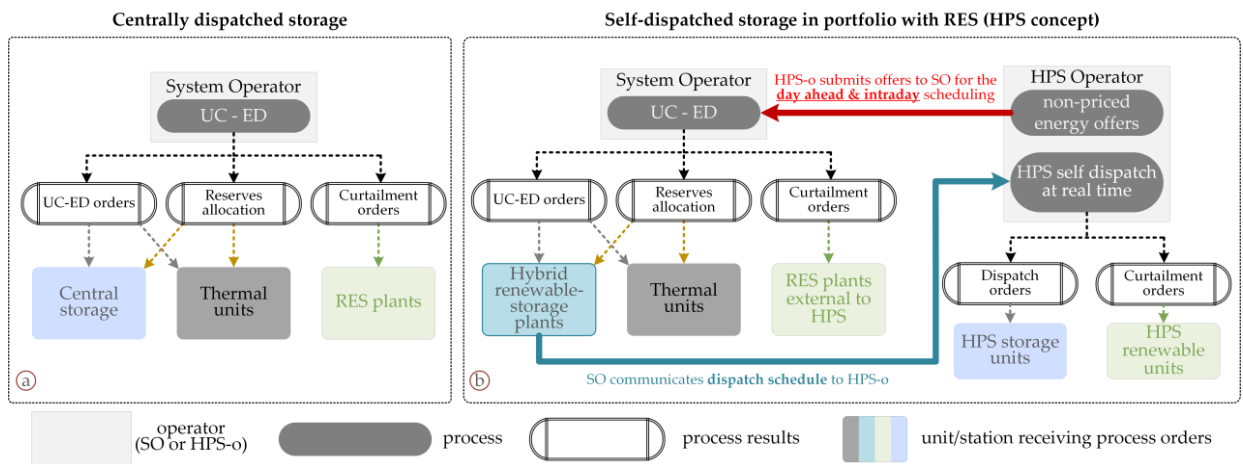

Fig. 1. High-level system management for (a) centrally-dispatched and (b) self-dispatched storage.

*2.3. Self-dispatched storage management*

Under the HPS concept, storage and renewable generation are integrated into a virtual power plant, to participate as a single dispatchable entity in the island market. The HPS and its constituents are managed by their operator (the HPS-o), interacting with the island system and market operator (the SO) to submit offers and receive dispatch instructions (Fig. 1(b)).

To participate in the day-ahead market, the HPS-o submits non-priced energy offers to the SO, covering the entire 24-h horizon of the dispatch day. HPS energy offers are translated into an hourly dispatch schedule through the day-ahead scheduling (DAS) solved by the SO, enjoying priority against conventional dispatchable generation. DAS results, in the form of hourly dispatch schedules for energy and reserves, are issued by the SO to all participants, including the HPS-o, who is then responsible to coordinate and self-dispatch the BES and renewable components of the HPS to comply with the dispatch orders of the SO, else imbalances will occur during real-time operation, incurring additional costs and penalties.

To meet generation dispatch orders, the HPS will use available renewable production, with the storage facilities complementing and balancing its variability. More specifically, during real-time operation the HPS may self-dispatch its components in the following modes:

- Available renewable production from the HPS RES units is stored in the HPS storage facilities.
- Available renewable production from the HPS RES units is directly injected into the island grid to fulfil the energy dispatch schedule issued by the SO.
- Energy previously stored in the HPS storage facilities is directly injected to the island grid to fulfil the energy dispatch schedule issued by the SO.
- Energy is absorbed from the island grid and stored in the HPS storage facilities. For the HPS to operate in this mode, the SO should issue a dispatch order for absorption from the grid.

An intra-day revision cycle of the UC-ED is foreseen at mid-day, when the HPS-o may update the submitted energy offer, to correct its position for the second half of the dispatch day and better exploit available renewable energy. Energy offer updates concern only the second half of the day (last 12-h interval).

Apparently, in this storage management paradigm, the BES is not a system-level tool at the disposal of the SO, but an asset belonging to the portfolio of a specific market participant.

## 3. Study Case & Assumptions

A medium-sized island with high RES penetration is used as the study case. Peak demand is ~210 MW, with a load factor of 46%. Conventional generation comprises 18 units whose technoeconomic characteristics can be found in [16]. Installed capacity of wind farms (WFs) and photovoltaics (PVs) in the baseline scenario without storage is 55 MW and 36 MW respectively, leading to annual RES penetration of 22.9%, with 12.5% RES energy curtailments. The capacity factors of WFs and PVs are ~40% and ~21% respectively, indicative of the good RES potential of the island. For the economic analysis, an oil price of 90 €/bbl and a cost of 40 €/tn for $CO_2$ emission rights are assumed.

To evaluate performance under substantially higher RES penetration levels, an additional WF

Table 1: BES and Wind Farm Investment Costs.

| BES investment cost | | | WF investment cost [€/kW] |
|---|---|---|---|
| Energy component [€/kWh] | | Power component [€/kW] | |
| Initial | Replace | | |
| 250 | 150 | 400 | 1200 |

Table 2: LCOE Calculation Parameters.

| Evaluation period (y) | Tax rate (TR) | O&M cost (OM) | Depreciation (DP) | Discount rate (i) |
|---|---|---|---|---|
| 20 years | 25% | 2% of initial investment | Linear, 10 years | 8% |

capacity of 75 MW is assumed, which comes along with BES storage, deployed in either of the two concepts discussed so far, to balance the increased RES intermittency and manage curtailments. Overall, the system hosts a total of 130 MW WFs and 36 MW PVs.

For the self-dispatch management concept, HPS-BES capacities ranging from 30 to 70 MW with a 10 MW step are considered; BES energy capacity varies in each case from 6 to 15 equivalent hours of operation at HPS maximum power. For the centrally dispatched storage concept, BES configurations are examined with a rated power varying from 7.5 MW to 75 MW with a 7.5 MW step, and storage capacities of 1, 2.5, 5, 7.5 and 10 equivalent hours of operation at full power. BES and RES investment costs are presented in Table 1. An 80% BES roundtrip efficiency is assumed in all cases. Batteries of li-ion technology are assumed in this paper. The parameters required for economic analysis and LCOE calculation are presented in Table 2.

Network-related constraints are not accounted in system modelling, as the focus of the paper is on the generation system of the island and on the benefits anticipated by the introduction of storage under the central or self-dispatching concept. Nevertheless, network congestion may exist in specific cases, imposing further constraints and increased renewable production curtailments. In such cases, proper siting of BES stations in the network becomes significant ([45,46]), while the exploitation of the dynamic thermal rating of transmission lines ([47]) can provide mitigation. Network congestion can be introduced in the analysis via techniques allowing the incorporation of linearized AC power flow constraints and network topologies in the MILP problem ([48]).

## 4. Modelling and Scenario Evaluation Methodology

Fig. 2 presents the simulation procedure employed. Each scenario is flagged "1" if it involves centrally managed storage or "2" if incorporating an HPS, in order to save the annual operating results of the system in the correct database.

To simulate the operation of the island system, the three-layer MILP-based optimization problem shown in Fig. 2 is employed. The first layer consists of two UC-ED generation scheduling processes, performed by the SO; the second and third layers refer to the real-time HPS self-dispatch (by the HPS-o) and the system economic dispatch (by the SO).

For each dispatch day, *D*, a DAS UC-ED optimization problem with a 24-h look-ahead horizon is initially solved by the SO, aiming at minimizing the overall system variable operating cost and maximizing RES penetration. DAS takes into account the energy offers of the HPSs, RES forecasts and the techno-economic characteristics of all units, to determine the commitment status and hourly load dispatch for all dispatchable assets. An intra-day UC-ED stage is foreseen at mid-day (T=12), where the optimization problem is re-executed with a 12-h look-ahead horizon, for the second half the day, based on updated energy offers from the HPS.

During real time operation, the HPS-o solves an internal self-dispatch optimization problem, subject to the dispatch orders issued by the SO, to maximize revenues. The SO performs the real-time dispatch of the system in the third layer, for the given commitment status of all units, to optimize system operation and balance any deviations.

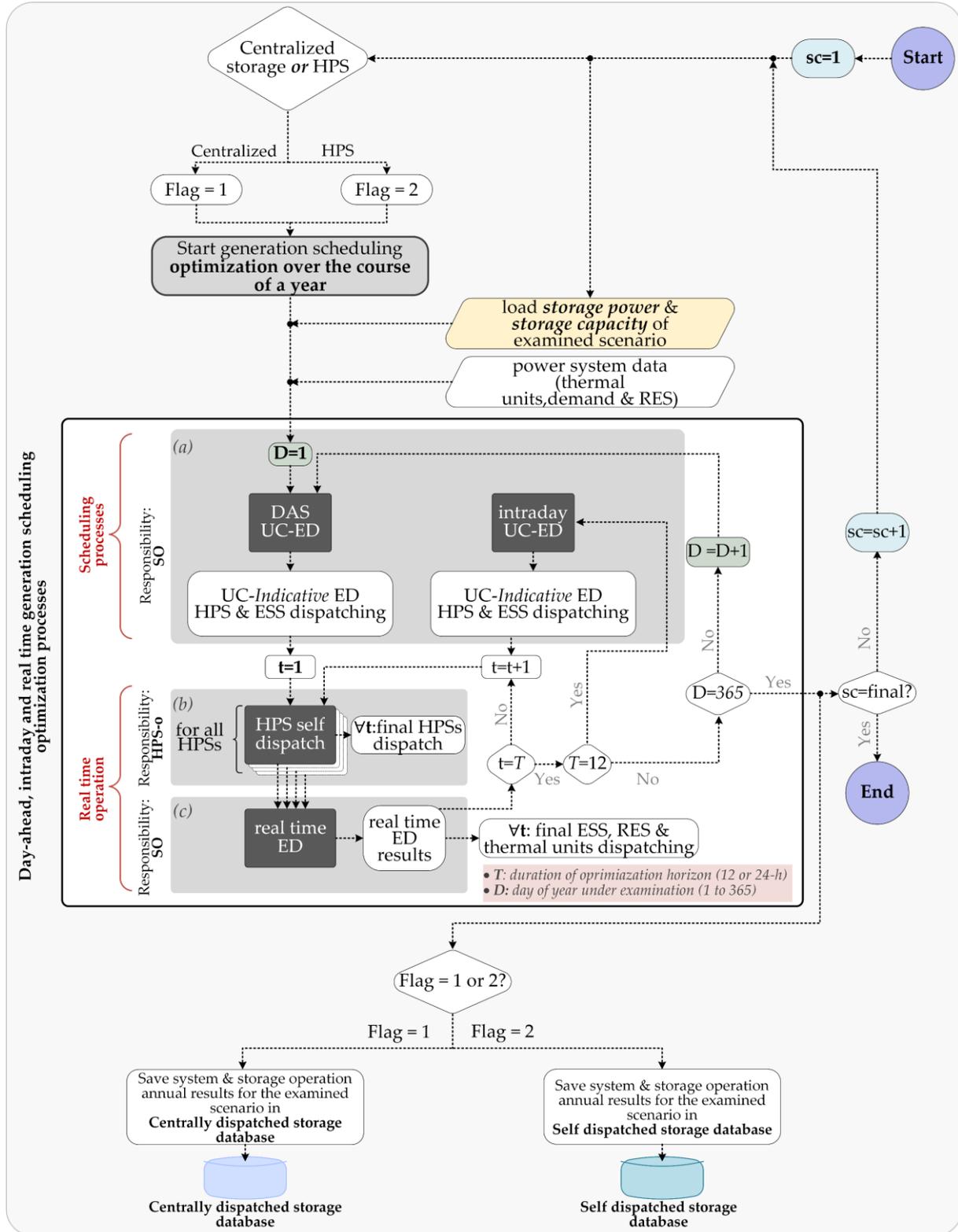

Fig. 2. Flow-chart of the entire simulation process. In the central frame, the three-layer daily generation scheduling process is outlined, repeated over the course of a year. It consists of: (a) System-level day-ahead (DAS) and intraday UCED, performed by the SO; (b) HPS self-dispatch, performed by the HPS operator; (c) system real-time dispatch and operation, by the SO.

To conduct annual simulations of the island system operation, this process is repeated for the 365 days of the year. The MILP optimization problem is implemented in GAMS [49], using the CPLEX optimizer [50].

As mentioned, comparison of central and self-dispatch storage deployment paradigms is based on the economic evaluation of the combined investment in new renewable capacity and BES facilities. In both cases, the metric used to evaluate the economic merit of each approach is the LCOE index, expressing the aggregate generation cost of the additional wind capacity (75 MW in all cases) and the storage facilities considered in each scenario. These metrics are *ex-post* calculated by retrieving the necessary data of each scenario from the respective database of Fig. 2.

*4.1. Generation Scheduling optimization*

*4.1.1. DAS and intraday UC-ED modelling*

The following MILP-based mathematical formulation covers both the DAS and the intra-day UC-ED processes, their main difference being the optimization horizon, respectively 24-h and 12-h. The objective function of the problem, (1), aims at minimizing the variable operating costs of conventional units ($C_p$), their start up ($C_{su}$) and shut-down costs ($C_{sd}$), the cost of reserves ($C_e$) and a cost for energy absorbed from the grid by HPSs ($C_{hps}^{gr}$). Cost terms in (1) are further analyzed in (2) to (7). Eq. (2) includes the linearized cost function of the online conventional units according to the methodology of [51]. The rightmost term of (1), $C_{sl}$, denotes the cost of slack variables allowing the relaxation of constraints to ensure feasibility of the solution and is calculated by (7). Slack variables are all those relaxing the active power and reserves equilibria, with the penalty factor attributed to each variable being related to the significance of the respective constraint (see (8)). For instance, the penalty factor assigned to the active power equilibrium slack variable is much higher than the respective penalty factors assigned to the active power reserves equilibria slack variables.

$$\min\left\{C_p + C_{su} + C_{sd} + C_e + C_{hps}^{gr} + C_{sl}\right\} \tag{1}$$

$$C_p = \sum_t\left(\sum_u\left(c_u^{p^{min}} \cdot st_{u,t} + \sum_b\left(\vartheta_{u,b} \cdot \Delta P_{u,t,b}\right)\right)\right) \tag{2}$$

$$C_{su} = \sum_t\sum_u c_u^{SU} \cdot su_{u,t} \tag{3}$$

$$C_{sd} = \sum_t\sum_u c_u^{SD} \cdot sd_{u,t} \tag{4}$$

$$C_e = \sum_t\sum_e\left(\sum_u c^e \cdot r_{u,e,t} + \sum_\xi c^e \cdot r_{\xi,e,t}\right) \tag{5}$$

$$C_{hps}^{gr} = \sum_t\sum_h f_{hps}^{gr} \cdot P_{h,t}^{gr} \tag{6}$$

$$C_{sl} = \sum_t\sum_e f_e \cdot x_{e,t} + \sum_t f_{ens} \cdot P_{ens,t} \tag{7}$$

$$f_e \ll f_{ens} \qquad (8)$$

Constraint (9) is the active power equilibrium of the system: in any hour $t$, the production of dispatchable units, $P_u$, BESs, $P_{\xi}^d$, wind ($P_w$) and photovoltaics ($P_{pv}$), excluding curtailments ($x_w$), and the amount of energy not served ($P_{ens}$) should equal demand, $P_l$, BESs charge, $P^c$, and HPS grid absorption, $P_h^{gr}$. Note that HPSs are included in the set $\mathcal{U}$ of dispatchable units. Constraint (10) enforces the fulfilment of reserves requirements, $rr$, per reserve type, $e$. It applies for three types of reserves: primary ($pr$), secondary ($sr$) and tertiary ($tr$). Reserves, $r$, are provided by conventional units, HPSs and BESs. Eq. (11) quantifies the primary up (*positive*) reserves requirements of the system, which are most important in isolated grids. A detailed discussion for the quantification of reserves requirements in island systems can be found in [14].

$$\sum_{u \in \mathcal{U}} P_{u,t} + \sum_{\xi} P_{\xi,t}^d + P_{pv,t} + P_{w,t} - x_{w,t} + P_{ens,t} = P_{l,t} + \sum_{\xi} P_{\xi,t}^c + \sum_{h} P_{h,t}^{gr} \qquad (9)$$

$$\sum_{u} r_{u,e,t} + \sum_{\xi} r_{\xi,e,t} + x_{e,t} \geq rr_{t,e}, \quad \forall e \in \{pr, sr, tr\} \qquad (10)$$

$$rr_{t,pr}^{up} = \max\left(\max_{u}\left(P_{u,t}\right), P_{w,t} - x_{w,t}\right) \qquad (11)$$

The fundamental technical constraints for the management of dispatchable units are (12) to (15). Constraints (12) and (13) impose the logical status of unit commitment, $st$, using binary variables defining the start-up, $su$, and shut-down, $sd$, of each unit. Constraint (14) enforces the ramp rates ($ru$, $rd$) applicable for each unit, while (15) and (16) bound the unit's output by its maximum ($P^{max}$) and minimum ($P^{min}$) power. Minimum up and down time constraints are also incorporated in the problem by (17) and (18) respectively.

$$su_{u,t} + sd_{u,t} \leq 1 \qquad (12)$$

$$su_{u,t} - sd_{u,t} = st_{u,t} - st_{u,t-1} \qquad (13)$$

$$-ru_u \cdot st_{u,t} \leq P_{u,t-1} - P_{u,t} \leq rd_u \cdot st_{u,t} + P_u^{\max} \cdot sd_{u,t} \qquad (14)$$

$$P_{u,t} + \sum_{e} r_{u,t,e}^{up} \leq P_u^{\max} \cdot st_{u,t} \qquad (15)$$

$$P_{u,t} - \sum_{e} r_{u,t,e}^{dn} \geq P_u^{\min} \cdot st_{u,t} \qquad (16)$$

$$\sum_{k=t-T_u^{run}+1}^{t} su_{u,k} \leq st_{u,t} \qquad (17)$$

$$\sum_{k=t-T_u^{stop}+1}^{t} sd_{u,k} \leq 1 - st_{u,t} \qquad (18)$$

The operation of each centralized BES system is governed by constraints (19) to (23). Eq. (19) sets the maximum charge and discharge rate ($P^{c/d\text{-}max}$), while (20) prevents the simultaneous charging and discharging using binary variables $st^{c/d}$. Constraints (21) and (22) determine the per hour state of charge ($SoC$) of the batteries, accounting for their roundtrip efficiency ($n$) and minimum and maximum acceptable operating levels ($E^{min/max}$). Constraint (23) defines the contribution of BES in the provision of positive reserves.

$$0 \leq P_{\xi,t}^{c/d} \leq P_{\xi}^{c/d-max} \cdot st_{\xi,t}^{c/d} \tag{19}$$

$$st_{\xi,t}^{c} + st_{\xi,t}^{d} \leq 1 \tag{20}$$

$$SoC_{\xi,t} = SoC_{\xi,t-1} + P_{\xi,t}^{c} \cdot \sqrt{n_{\xi}} - P_{\xi,t}^{d}/\sqrt{n_{\xi}} \tag{21}$$

$$E_{\xi}^{min} \leq SoC_{\xi,t} \leq E_{\xi}^{max} \tag{22}$$

$$P_{\xi,t}^{d} + \sum_{e} r_{\xi,r,e}^{up} \leq P_{\xi}^{d-max} + P_{\xi,t}^{c} \tag{23}$$

The operation of HPS from a SO perspective is subject to constraints (24) to (30). In eq. (24), the aggregate dispatch order, $P_h$, issued in hour $t$ to an HPS, $h$, is the sum of the UC-ED result for the individual controllable units of the HPS, limited by its maximum declared capacity, $P_h^{max}$. Constraint (25) defines the commitment status, $\ell$, of the HPS as a whole. Constraints (26) to (28) distribute the energy offer of the HPS, $E^{offer}$, in the dispatch day, taking account of the roundtrip efficiency, $c$, of the station. Constraint (29) defines the amount of HPS non-dispatched energy, $x_h$. Finally, (30) delimits absorption from the grid to remain below $P_h^{gr-max}$ and prevents simultaneous energy production and absorption by the HPS.

$$P_{h,t} = \sum_{u} P_{u,t} \leq P_h^{\max}, \quad \forall h \mid u \in \mathcal{U}_h \tag{24}$$

$$\ell_{h,t} \cdot \min_{u} \{P_u^{\min}\} \leq P_{h,t} \leq \ell_{h,t} \cdot \sum_{u} P_{u,t}^{\max}, \quad \forall h \mid u \in \mathcal{U}_h \tag{25}$$

$$E_{h,t}^{av} \leq E_{h,t-1}^{av} + c_h \cdot P_{h,t-1}^{gr} - P_{h,t-1}, \forall t \geq 2 \tag{26}$$

$$E_{h,1}^{av} \leq E_h^{offer} \tag{27}$$

$$P_{h,t} \leq E_{h,t}^{av} \tag{28}$$

$$x_h = E_h^{offer} + c_h \cdot \sum_{t} P_{h,t}^{gr} - \sum_{t} P_{h,t} \tag{29}$$

$$P_{h,t}^{gr} \leq (1 - \ell_{h,t}) \cdot P_{h,t}^{gr-\max} \tag{30}$$

To determine the HPS energy offer $E^{offer}$, only a few strategies are available in the literature, [34,52,53]. Simple approaches ([34,52]) rely on the application of safety coefficients on the forecasted HPS-RES energy, to avoid over-optimistic offers that could eventually lead to real-time imbalances for the station. Other, more sophisticated approaches use optimization techniques to build the energy offer of the HPS, employing risk-aversion methodologies, such as in [53]. In this paper, the safety coefficients method is used to construct the HPS offers in the day-ahead and intraday UC-ED problems, with coefficients equal to 60% for the first 8-h, 50% for the following 8-h and 40% for forecasted RES production in the last 8-h.

### 4.1.2. HPS self-dispatch during real time operation modelling

The main target of the HPS-o during real-time operation is revenue maximization while complying with dispatch orders of the SO to avoid imbalances. This is formulated as a static MILP problem, simplifying the methodology of [41]. The objective function (31) includes revenues from energy sales to the grid (1st term), expenses for purchase of energy from the grid and imbalance charges (2nd and 3rd terms). Parameters $m_{1/2/3}$ denote unit prices for energy sales, purchases and imbalance penalties. The time dimension is neglected due to the static nature of the problem. Additionally, the $h$ index is also neglected in most variables of the following constraints.

$$\max\left\{m_1 \cdot \left(P_{hps}^{res-g} + P^{dch}\right) - m_2 \cdot P_h^{gr} - m_3 \cdot \left(P_{hps}^{imb-p} + P_{hps}^{imb-a}\right)\right\} \quad (31)$$

HPS technical constraints are introduced in (32)-(37). Eq. (32) expresses that available energy by the HPS RES units ($P_{hps}^{res}$) can be either directly supplied to the grid ($P_{hps}^{res-g}$), in partial or full fulfillment of the HPS dispatch order for production, stored in the HPS storage facilities ($P_{hps}^{res-s}$) or rejected ($P_{hps}^{res-r}$). HPS dispatch orders for production are fulfilled in combination by RES energy directly injected to the grid and by HPS storage discharge, $P^{dch}$, as in (33), where production imbalances, $P_{hps}^{imb-p}$, are foreseen. Eq. (34) refers to charging ($P^{ch}$) of the HPS storage systems, via grid absorption, based on respective dispatch orders issued by the SO, and allowing for RES recharging, $P_{hps}^{res-s}$, and imbalances, $P_{hps}^{imb-a}$. The energy levels in the HPS storage facilities ($E^{hps}$) vary subject to (35) and (36), while (37) defines the maximum charging and discharging capability of the HPS units ($P^{ch/dch-max}$).

$$P_{hps}^{res-s} + P_{hps}^{res-g} + P_{hps}^{res-r} = P_{hps}^{res} \quad (32)$$

$$P_{hps}^{res-g} + P^{dch} + P_{hps}^{imb-p} = P_h \quad (33)$$

$$P_h^{gr} + P_{hps}^{imb-a} = P^{ch} - P_{hps}^{res-s} \quad (34)$$

$$E^{hps} - E_0^{hps} = \sqrt{c} \cdot P^{ch} - P^{dch}/\sqrt{c} \quad (35)$$

$$E_{min}^{hps} \leq E^{hps} \leq E_{max}^{hps} \quad (36)$$

$$0 \leq P^{ch/dch} \leq P^{ch/dch-max} \quad (37)$$

### 4.1.3. System real time economic dispatch modelling

The real-time ED (RT-ED) of the system is simulated based on a simple MILP approach, assuming the commitment status of conventional units as determined in prior generation scheduling stages. The objective function (38) is similar to (1), excluding the start-up and shut-down costs of conventional units, which are not relevant here.

$$\min\left\{C_p + C_e + C_{hps}^{gr} + C_{sl}\right\} \quad (38)$$

The problem is subject to (9)-(11), (14)-(16) and (19)-(30), excluding the time dimension. HPS energy offers do not exist specifically for each RT-ED interval. Hence, in (27), energy offers are substituted by the dispatch schedule determined during UC-ED for the respective time interval, augmented by the amount of non-dispatched HPS energy in the previous hours up to the examined time period. During real-time operation, the secondary and tertiary reserves allocated in the previous stage may be released, thus allowing the RT-ED static optimization algorithm to maximize exploitation of available RES potential.

Centrally managed BES stations, on the other hand, are flexible system assets, not bound by offers and revenue maximization concerns. Hence, their charging and discharging operation can be rescheduled during RT-ED to absorb renewable energy that would otherwise be curtailed. However, as RT-ED is only a static optimization process, it cannot serve energy arbitrage functionality. For this reason, the BES reference SoC profile derived in the DAS and intraday UC-ED processes is maintained as a minimum SoC requirement in real-time optimization to ensure availability of energy reserves for the battery station to perform the intended arbitrage function on the daily load curve.

*4.2. LCOE calculation*

LCOE formulae ((39) and (40)) have been built upon the classic approach proposed in [54], that takes into account the income tax rate and depreciation of the investments.

The LCOE of the aggregate investment in renewables and centralized storage is calculated according to (39), where $E^{res}$ represents the energy injected to the grid by the additional wind capacity introduced to the system, which constitutes an objective measure of the energy contribution of the combined RES and storage facilities, to establish an equal basis of comparison with the HPS case. Note that $E^{res}$ corresponds to the annual output of the additional wind capacity (75 MW), excluding energy transactions of the BES, which is operated to serve diverse functionalities, not relevant in this calculation (e.g. provision of reserves, arbitrage etc.). The discharge of the centrally dispatched BES is not accounted for in (39) to avoid double counting RES production of the new wind farms that is absorbed and then reinjected into the grid. BES may also be charging by absorbing energy from other sources in the grid, conventional or RES, distorting the calculation of the LCOE for the new assets.

$$LCOE^c = \frac{I_0^{bes+res} + \dfrac{I_{10}^{bes}}{(1+i)^{10}} + \sum_{y=1}^{20} \dfrac{(OM_y) \cdot (1-TR_y) - DP_y \cdot TR_y}{(1+i)^y}}{(1-TR_y) \cdot \sum_{y=1}^{20} \left( \dfrac{E_y^{res}}{(1+i)^y} \right)} \quad (39)$$

For the self-dispatched HPS concept, the LCOE is given by (40), where the energy used in the denominator is the net energy production of the hybrid station, i.e. the output of its RES units reaching the network, either directly or through the storage facilities. In this case, BES are only charging and discharging wind energy generated by the HPS-WFs. Thus, the calculation of the LCOE in this case in not compromised by double counting the same amount of energy, as in (39). HPS imbalance charges are also accounted for in (40).

$$LCOE^s = \frac{I_0^{hps} + \dfrac{I_{10}^{bes}}{(1+i)^{10}} + \sum_{y=1}^{20} \dfrac{\left(OM_y + \left(P_y^{imb-p} + P_y^{imb-a}\right) \cdot m_3\right) \cdot (1-TR_y) - DP_y \cdot TR_y}{(1+i)^y}}{(1-TR_y) \cdot \sum_{y=1}^{20} \left( \dfrac{\sum_{t=1}^{8760}\left(P_{hps}^{res-g} + P^{dch}\right)}{(1+i)^y} \right)} \quad (40)$$

In the LCOE calculations, one replacement of the batteries is foreseen after 10 years of operation, at the replacement cost given in Table 1. BES replacement becomes necessary due to the loss of life predominantly induced by cycle aging, [55,56].

## 5. Results & Discussion

*5.1. Daily system operation*

Fig. 3 presents the operation of the island system for an indicative medium-demand week, when 75 MW of new WFs have been installed in the system. This new WF capacity is either incorporated in a 30 MW HPS, comprising 240 MWh of BES facilities, or is supported by a 30 MW/240 MWh BES station centrally managed by the SO. Both storage management concepts lead to similar levels of renewable energy curtailments during the examined week, with the centralized approach enjoying a slight advantage. The increased curtailments in the first 36-h period in Fig. 3 are due to the prevailing high wind and low demand conditions.

Despite similarities in operation from a system viewpoint, BES facilities operate in a substantially different manner, as shown by the weekly SoC variation profiles in Fig. 4. In the HPS case, battery charging tracks the variation of HPS wind production, while discharging is subject to the dispatch orders issued by the SO. Hence, the HPS batteries in Fig. 4(a) are often fully charged, especially in periods with increased wind availability, storing energy for exploitation in subsequent dispatch periods.

The SoC of centrally managed BES facilities (Fig. 4(b)) follows a notably different pattern, remaining at low charge levels, as the batteries in this concept facilitate direct injection of RES energy in the system by providing increased levels of primary up reserves to relax RES absorption constraints ([16]), rather than storing excess RES energy for use in subsequent dispatch periods.

*5.2. Annual operating results*

Fig. 5 and Fig. 6 present annual operating results for the self-dispatch concept and Fig. 7 for the centrally dispatched storage. In all cases, the same additional WF capacity is introduced into the system (75 MW). Note that in Fig. 6, wind energy curtailments are differentiated for wind farms external (existing 55 MW) and internal to the HPS (new 75 MW). External WFs are managed via setpoint commands issued by the SO for technical and security reasons, whilst the HPS WFs are subject to the internal management strategy adopted by the HPS-o. In the centralized case (Fig. 7(b)), wind curtailments refer to the entire WF capacity of the system (130 MW in total), all managed by the SO.

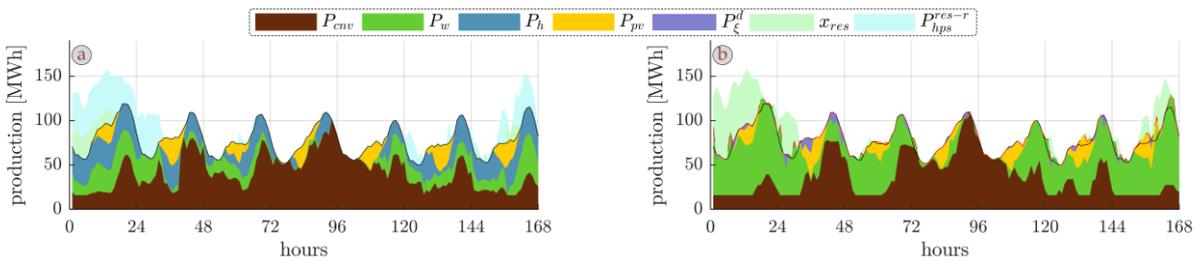

Fig. 3. System operation during a typical week, with the addition of 75 MW of WFs supported by (a) self-dispatched and (b) centrally dispatched BES. $P_{cnv}$ refers to the cumulative production of online conventional units.

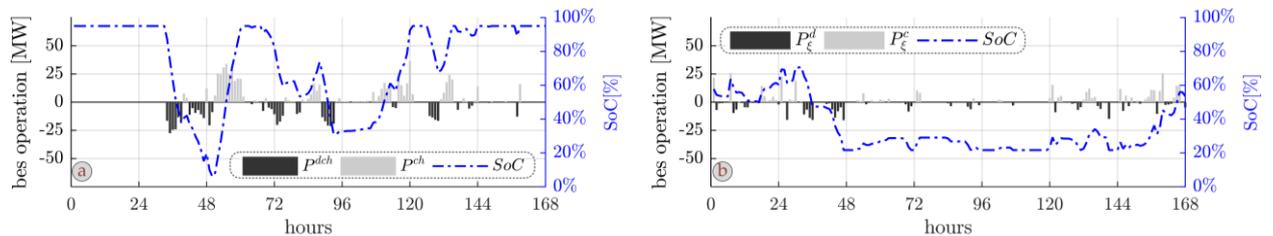

Fig. 4. SoC variation of (a) self-dispatched and (b) centrally dispatched BES for the same week as in Fig. 3.

The contribution of storage in enhancing RES penetration is evident in Fig. 5 (a) and Fig. 7(a), for the self and central dispatch management concept respectively. The achieved RES penetration in both cases ranges between 40% to 50%, depending on storage capacity available, however there do exist substantial differences as discussed in the following.

In the self-dispatch concept, the assumed rated power $P_h^{max}$ of the HPS does not play an essential role in achieving high RES penetrations; it is rather the energy capacity of storage that leads to high penetration rates, as time-shifting of available RES production plays a central role in the HPS concept.

In the case of centralized storage, on the other hand, energy capacity plays a secondary role, with the power rating of BES facilities proving to be instrumental in achieving increased RES penetrations. For example, a 45 MW BES with a capacity of only 90 MWh (2 h) will lead to 47.3% annual RES penetration, while expanding its capacity to 450 MWh will only increase RES penetration to 47.9%. This happens because the centralized BES systems primarily provide fast operating reserves to relax RES penetration constraints, rather than time-shift wind production as in the case of the HPS. In this task, it is the power rating of the batteries, rather than their energy capacity, that plays a central role.

Increased RES penetration always comes at the expense of curtailments of available wind energy. In the self-dispatch storage concept, to achieve HPS-WF curtailments below 20%, a BES capacity of ~600 MWh is required (red squares in Fig. 6). At the same time, existing WFs (blue triangles in Fig. 6) experience increased rejections as the level of congestion in the system increases, eventually reaching similar levels as for the HPS WF for this particular storage sizing. In the centralized concept, the same overall level of wind curtailments can be attained with batteries of small storage capacities even below 100 MWh (Fig. 7(b)).

The computed LCOE values reveal an enormous gap between the two BES management concepts. The LCOE of the centrally dispatched BES variant starts from ~80 €/MWh for small facilities; the respective LCOE of the cheapest self-dispatched HPS variant is almost double, beginning from ~160 €/MWh. Differences are due to the high energy capacity requirements of the HPS paradigm, which impacts the investment cost of the solution. The gap between the two concepts remains large even when storages of similar sizes are examined. For instance, when a 400 MWh BES is considered, the lowest LCOE obtained for the self-dispatched HPS concept is ~177 €/MWh, against ~143 €/MWh for the centralized case. With the investment cost of the two options being the same in this example, the difference in LCOE is due to the most efficient exploitation of available renewable energy achieved by the centralized storage.

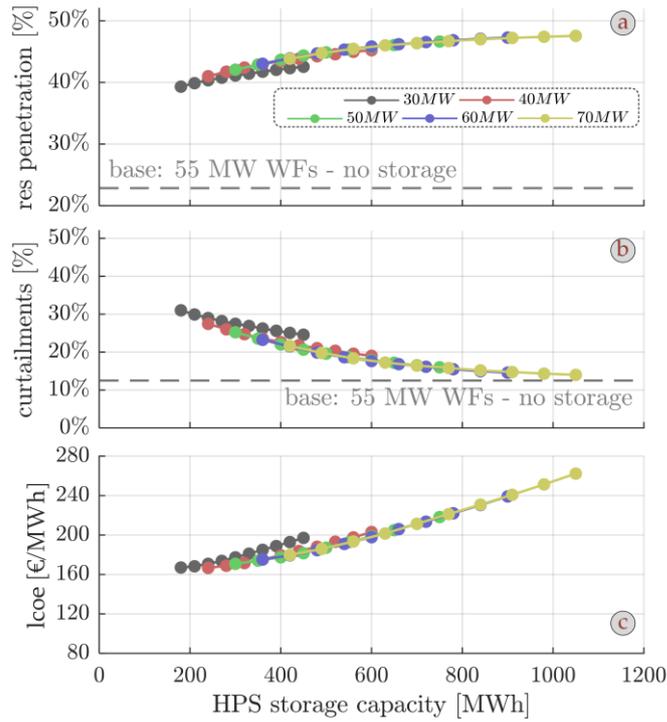

Fig. 5. Annual operating results for the self-dispatch (HPS) concept, with 75 MW of additional wind capacity. (a) RES penetration levels, (b) curtailments for the entire WF capacity of the system (internal and external to the HPS, 55+75 MW) and (c) LCOE of the combined investment in new wind and BES. All plotted against installed BES energy capacity; different colored dots indicate HPS power capacities.

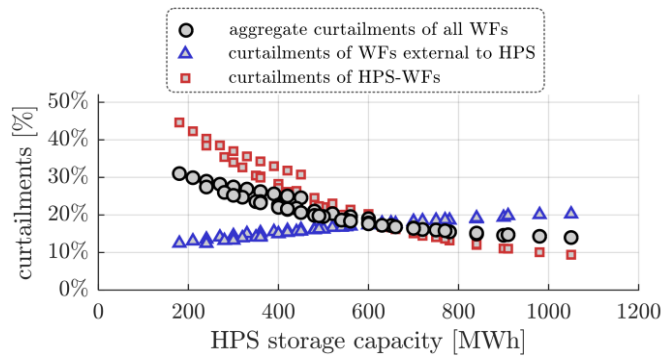

Fig. 6. Annual wind curtailments: in red squares for wind farms belonging to the HPS (75 MW), in blue triangles for existing WFs external to the HPS (55 MW) and in black circles for all WFs of the system (55+75 MW).

### 5.3. Pareto optimality of LCOE vs. RES penetration

From Fig. 5 and Fig. 7 it is evident that for a given RES portfolio, multiple BES configurations can lead to similar RES penetration levels, yet at a significantly different LCOE. Evidently, the maximization of system RES penetration levels and the minimization of combined storage-renewables LCOE are conflicting objectives, with the latter not being directly accounted for in the simulation process. The Pareto fronts in Fig. 8 show the storage sizing that maximize RES penetration at minimum cost. Note that the Pareto fronts of Fig. 8 are implicitly calculated from the annual RES penetration results of each examined scenario along with the *ex-post* calculation of the LCOE of the respective configuration and design. The sizing of all BES configurations belonging to the Pareto fronts is further analyzed in Fig. 9.

From Fig. 8 it is apparent that a certain RES penetration can be reached in a significantly more economical manner employing the centralized storage concept. For instance, annual penetrations around 46% could be achieved via a combined investment in 75 MW of wind and 37.5MW/37.5MWh of centralized BES, that will require a remuneration lower than 80 €/MWh to ensure feasibility; the same penetration could be also reached by a HPS with the same amount of installed wind capacity but a much higher battery capacity of 660 MWh, that would require a compensation of ~200 €/MWh.

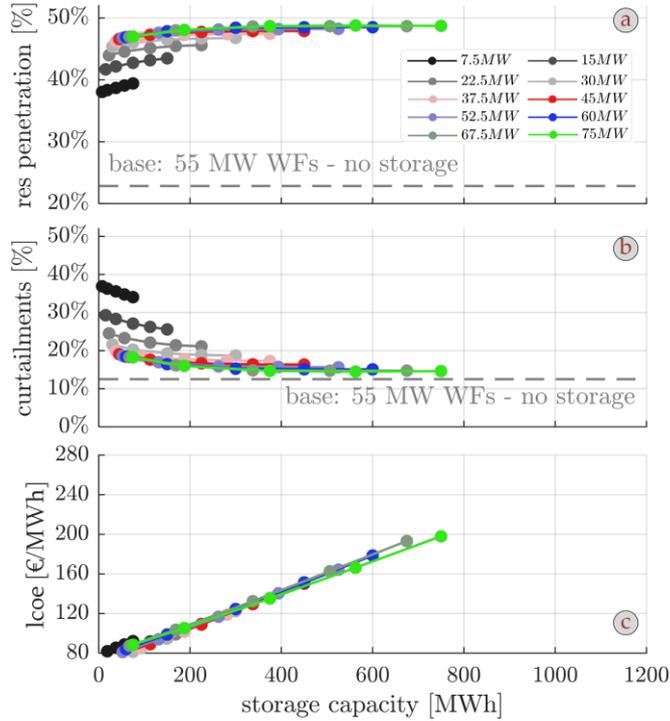

Fig. 7. Annual operating results for the centralized storage concept, with 75 MW of additional wind capacity. (a) RES penetration levels, (b) curtailments for the entire WF capacity on the island (55 MW+75 MW) and (c) LCOE of the combined investment in new wind and BES. All plotted against installed BES capacity; different colored dots indicate BES power capacities.

As already discussed, these differences are due to the fundamentally different exploitation of storage in the examined management concepts. In the centralized case, the battery is a system flexibility asset at the disposal of the SO, allowing full utilization of the entire range of available services to optimize system operation and RES penetration. Hence, storage primarily provides fast response reserves, relaxing RES penetration constraints of the system. On the other hand, the self-dispatch HPS concept aims at transforming intermittent renewables into dispatchable generation, via a VPP-type solution where the batteries become an internal asset of the HPS, used for balancing wind variability and providing firm capacity, while at the same time performing extensive time shifting of renewable energy to conform to a dispatch schedule issued by the SO, eventually leading to substantially larger capacity requirements and cost.

*5.4. Impact on system economics*

In Fig. 10, the impact on the annual generation cost of the island system is calculated for each BES configuration lying on the Pareto front of Fig. 8, with respect to the base case scenario without additional renewables and storage. The cost of each scenario includes the variable cost of all conventional units, the remuneration of renewable production participating in the energy mix at predefined tariffs, equal to 65 €/MWh for WFs and PVs already existing on the island, while the combined new investments in additional WFs and BES are remunerated at the respective LCOE (calculated based on (39) and (40)). Negative values in Fig. 10 denote an increase in system cost, while positive differences correspond to savings due to the substitution of more expensive conventional energy by RES, as well as to the optimization of the overall generation system management afforded by the introduction of flexible assets.

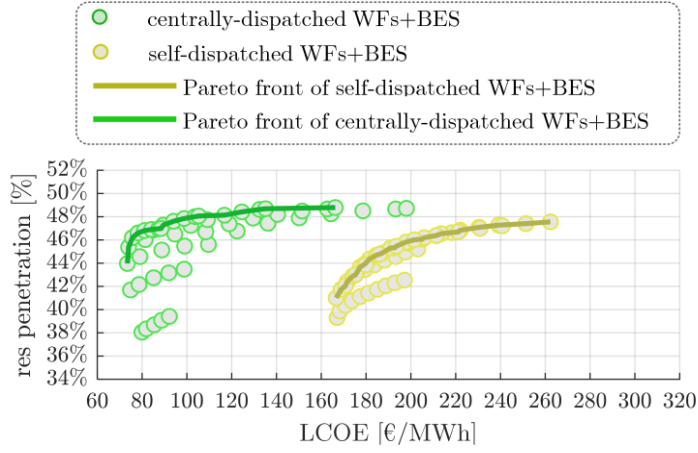

Fig. 8. Pareto front of RES penetration vs LCOE of all wind-BES configurations evaluated, for the central and self-dispatch storage concepts.

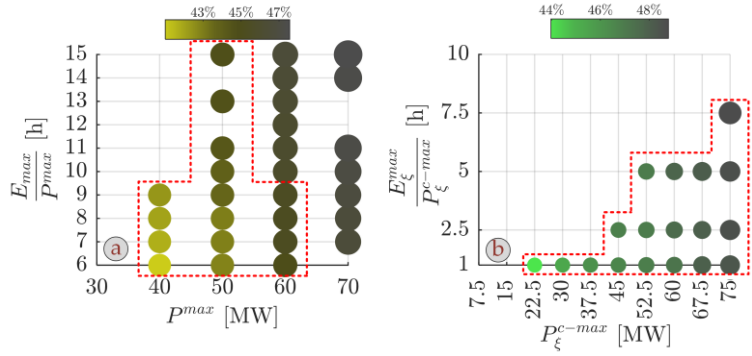

Fig. 9. Power and energy capacities of all BES configurations on the Pareto fronts of Fig. 8. (a) Self-dispatch and (b) central dispatch storage concept. Larger bubble sizes indicate a higher LCOE, darker bubble colors a higher RES penetration.

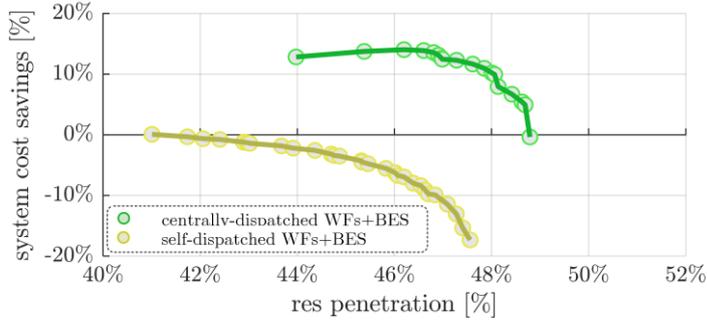

Fig. 10. Impact on annual variable generation cost of the system when the wind-BES configurations on the Pareto fronts of Fig. 8 are introduced into the system.

Evidently, savings in system cost are achieved by all centrally dispatched BES configurations. The self-dispatched HPS storage concept, being characterized by a higher LCOE, results in increased system cost for most Pareto optimal scenarios. Notably, only two HPS configurations (40MW/6h & 40MW/7h) have a neutral impact on system cost. For the particular island system, any BES-RES

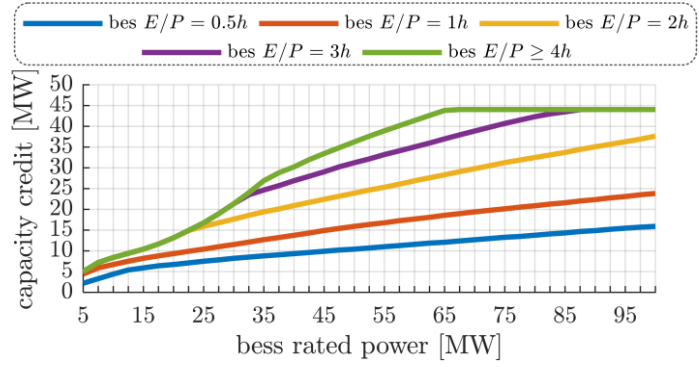

Fig. 11. BES contribution to resource adequacy for the study case power system, evaluated employing the load-levelling methodology.

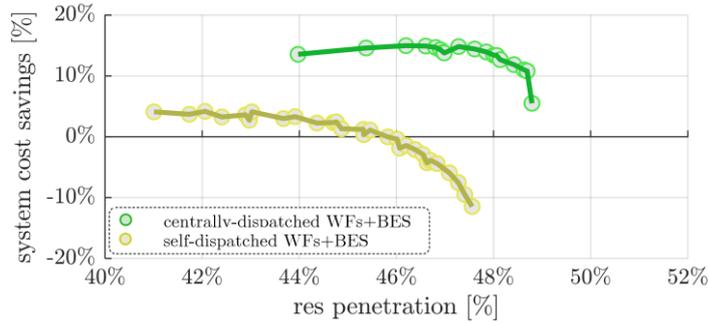

Fig. 12. Savings in annual total generation cost of the system achieved by wind-BES configurations on the Pareto fronts of Fig. 8. Value of storage contribution to resource adequacy taken into account, based on a firm capacity provision as in Fig. 11.

combination with an LCOE higher than ~165 €/MWh will impact negatively system economics.

However, cost variations presented in Fig. 10 ignore any potential contribution of storage to resource adequacy of the island system. Employing the load-levelling methodology ([57,58]) to derive an estimate of the contribution of storage to the capacity adequacy of the system, the results of Fig. 11 regarding the capacity credit of each storage configuration are obtained. It is observed that high energy capacity storages are more effective, while there is a limit to the potential contribution of storage through arbitrage, primarily related to the shape of the daily load curve of the system.

The capacity contribution of storage quantified in Fig. 11 signifies an avoided investment in equivalent thermal capacity, whose annual fixed cost can be considered an additional benefit to be credited to the respective storage solution. In this analysis, the investment cost of a thermal power station is taken equal to 1400 €/kW, yielding an annualized equivalent fixed cost of ~177 €/kW. Taking this into account, Fig. 12 depicts the impact on the total generation cost of the island system, including both the variable and fixed cost savings achieved through the introduction of BES configurations belonging to the Pareto fronts of Fig. 8. The self-dispatched wind-BES concept now proves to be cost-efficient for the system, as several HPS configurations (all enclosed by red dashed line in Fig. 9) lead to a reduction in system cost when remunerated at their respective LCOE.

Centrally dispatched BES are also characterized by enhanced system cost reductions in Fig. 12 compared to Fig. 10, however differences are smaller because their contribution to resource adequacy is limited by the low energy capacities of the Pareto optimal configurations. Nevertheless, with all factors considered, the centralized BES concept proves to be the most cost-efficient solution to support a high RES penetration in the island system.

# 6. Conclusions

In this paper a comparison is performed between the two alternative storage deployment and management concepts for island systems. A centrally dispatched solution, managed by the SO, is compared to a self-dispatched one, where storage is operated in unison with renewables under the HPS concept. Li-ion batteries are the study-case storage technology in the paper.

Island system management and operation, including RES and storage facilities, is modelled via a three-layer MILP-based process, representing day-ahead UC-ED and real time dispatch and operation. A large number of BES configurations is evaluated along with the introduction of new wind capacity, to determine which storage management concept leads to optimal results in terms of RES penetration and system economics.

The analysis indicates that both storage management concepts can support substantially increased RES penetration levels, over 45% on an annual basis. However, the centrally dispatched BES design proves to be more cost-efficient, as it requires a substantially smaller battery capacity compared to the self-dispatched HPS solution, leading to a reduced LCOE for the combined RES-storage investment required to achieve any specific RES penetration target and eventually managing to reduce the overall generation cost of the island system.

## CRediT authorship contribution statement

**G. N. Psarros:** Conceptualization, Resources, Methodology, Software, Validation, Formal analysis, Investigation, Data curation, Writing - original draft, Writing - review & editing, Visualization. **P. A. Dratsas:** Methodology, Software, Investigation, Writing - review & editing. **S.A. Papathanassiou:** Writing - review & editing, Supervision, Project administration, Funding acquisition.

## Declaration of competing interest

The authors declare that they have no known competing financial interests or personal relationships that could have appeared to influence the work reported in this paper.

## Acknowledgments

This work has been conducted in part within the Kythnos Smart Island project, awarded by the Siemens Electrotechnical Projects and Products Societe Anonyme.

## Nomenclature

*Abbreviations*

| | |
|---|---|
| *BES* | Battery energy storage |
| *DAS* | Day-ahead-scheduling |
| *DP* | Depreciation period |
| *FiT* | Feed-in-tariff |
| *HPS* | Hybrid power station |
| *HPS-o* | HPS operator/owner |
| *LCOE* | Levelized cost of energy |
| *MILP* | Mixed integer linear programming |
| *OM* | Operation & maintenance |
| *PVs* | Photovoltaics |
| *RES* | Renewable energy sources |
| *RT-ED* | Real time economic dispatch |
| *SO* | System Operator |
| *TR* | Tax rate |
| *UC-ED* | Unit commitment & economic dispatch |
| *VPP* | Virtual power plant |
| *WFs* | Wind farms |

*Sets*

| | |
|---|---|
| $\mathcal{B}$ | Set of indices for the piecewise linearization of the variable operating cost function of each conventional unit |
| $\mathcal{D}$ | Set of indices of days over the year |
| $\mathcal{E}$ | Set of indices of reserves types |
| $\mathcal{H}$ | Set of indices of HPS |
| $\mathcal{J}$ | Set of indices of centrally dispatched BES |
| $T$ | Set of indices of time intervals within optimization horizon |
| $\mathcal{U}$ | Set of indices of dispatchable units except for centrally dispatched BES |
| $\mathcal{U}_h \subset \mathcal{U}$ | Subset of indices of HPS dispatchable units |

*Indices*

| | |
|---|---|
| $t, k \in T$ | Time intervals of optimization horizon |
| $h \in \mathcal{H}$ | Hybrid power stations |
| $u \in \mathcal{U}$ | Dispatchable units |
| $\xi \in \mathcal{J}$ | Centrally dispatched BES |
| $e \in \mathcal{E}$ | Reserves type (*pr*-primary/*sr*-secondary/*tr*-tertiary) |
| $b \in \mathcal{B}$ | Number of blocks of the linearized cost function of each thermal unit |

*Binary variables*

| | |
|---|---|
| $sd_{u,t}$ | Binary variable equal to 1 if unit *u* shuts down at *t* |
| $st_{u,t}$ | Binary variable equal to 1 if unit *u* is online at *t* |
| $st^{c/d}_{\xi,t}$ | Binary variable equal to 1 if BES $\xi$ discharges (*d*) or charges (*c*) at *t* |
| $su_{u,t}$ | Binary variable equal to 1 if unit *u* starts up at *t* |
| $\ell_{h,t}$ | Binary variable equal to 1 if HPS *h* operates at *t* |

*Continuous variables*

| | |
|---|---|
| $C_p$ | Variable operating cost of conventional units over the optimization horizon |
| $C_{su/sd}$ | Start up and shut down cost of conventional units over the optimization horizon |
| $C_e$ | Cost of allocated reserves over the optimization horizon |
| $C^{gr}_{hps}$ | Cost of energy absorbed from the grid by HPS over the optimization horizon |
| $C_{sl}$ | Cost of slack variables |
| $c^{p^{min}}_u$ | Cost of operating at minimum loading for unit *u* |
| $c^{SU}_u$ | Start-up cost of unit *u* |
| $c^{SD}_u$ | Shut-down cost of unit *u* |
| $c^e$ | Cost of allocated reserves of type *e* |
| $E^{av}_{h,t}$ | HPS *h* available energy to be dispatched in period *t* |
| $E^{hps}$ | Energy stored in HPS storage facilities |
| $P_{u,t}$ | Production level of unit *u* at *t* |
| $P_{h,t}$ | Production level of HPS *h* at *t* |
| $P^{c/d}_{\xi,t}$ | Charging/discharging level of BES $\xi$ at *t* |
| $P_{ens,t}$ | Energy not served at *t* |
| $P^{gr}_{h,t}$ | Energy absorbed from the grid of HPS *h* at *t* |
| $P_{hps}^{res-s}$ | Available production by the HPS RES stored in the HPS storage facilities |
| $P_{hps}^{res-g}$ | Available production by the HPS RES directly supplied to the grid |
| $P_{hps}^{res-r}$ | Available production by the HPS RES rejected |
| $P^{dch}$ | HPS storage in discharge mode |
| $P^{ch}$ | HPS storage in charge mode |
| $P_{hps}^{imb-p}$ | HPS imbalance from production orders issued by SO |
| $P_{hps}^{imb-a}$ | HPS imbalance from grid absorption orders issued by SO |
| $r_{u/\xi,t,e}^{up/dn}$ | Upward (*up*) or downward (*dn*) reserves type *e* allocated to unit *u* or BES $\xi$ at *t* |
| $rr_{e,t}$ | System reserves requirements of type *e* at *t* |
| $SoC_{\xi,t}$ | State of charge of BES $\xi$ at *t* |
| $x_{e,t}$ | Slack variable denoting the violation of reserves requirements of type *e* at *t* |
| $x_h$ | The amount of HPS non-dispatched energy within the examined optimization horizon |
| $x_{w,t}$ | Wind curtailments at *t* of WFs external to HPS |
| $\Delta P_{u,t,b}$ | Production level of unit *u* at block *b* at *t* |

*Parameters*

| | |
|---|---|
| $c_h$ | Roundtrip efficiency of HPS *h* storage facilities |
| $E_h^{offer}$ | Energy offer of HPS *h* covering the entire optimization horizon (24-h or 12-h) |
| $E_\xi^{max/min}$ | Maximum/minimum state of charge of BES $\xi$ |
| $E_{max/min}^{hps}$ | Maximum/minimum state of charge of HPS storage facilities |
| $E_y^{res}$ | Energy injected into the grid by the additional wind capacity (75 MW) under the central-dispatch storage design |
| $f_e$ | Penalty factor for the violation of reserves requirements equilibrium *e* |
| $f_{ens}$ | Penalty factor for the violation of active power balance equilibrium |
| $f_{hps}^{gr}$ | Cost of energy absorbed from the grid by HPSs |
| $i$ | Discount rate |
| $I^{hps}$ | Investment in new HPS |
| $I^{bes}$ | Investment in new BES |
| $I^{bes+res}$ | Investment in new BES and RES |
| $LCOE^c$ | LCOE of the central-dispatch storage management concept |
| $LCOE^s$ | LCOE of the self-dispatch storage management concept |
| $m_{1-3}$ | Prices for energy sales, purchases and imbalance penalties of HPS |
| $n_\xi$ | Roundtrip efficiency of BES $\xi$ |
| $P_{pv,t}$ | Available energy produced by PV stations at *t* |
| $P_{w,t}$ | Available energy produced by the WFs external to HPS at *t* |
| $P_{l,t}$ | Load demand at *t* |
| $P_u^{max}$ | Maximum power output of unit *u* |
| $P_u^{min}$ | Minimum power output of unit *u* |
| $P_\xi^{c/d-max}$ | Max charging/discharging capacity of BES $\xi$ |
| $P_h^{\max}$ | Maximum declared capacity of HPS $h$ |
| $P_{h,t}^{gr-\max}$ | Maximum absorbing capability of HPS $h$ at $t$ |
| $P_{hps}^{res}$ | Available production by the HPS RES |
| $P^{ch/dch\text{-}max}$ | Max charging/discharging capacity of HPS storage facilities |
| $rd_u$ | Ramp down rate of unit $u$ |
| $ru_u$ | Ramp up rate of unit $u$ |
| $T_u^{run/stop}$ | Duration of minimum up/down time for unit $u$ |
| $y$ | Evaluation period |
| $\vartheta_{u,b}$ | Slope (marginal cost) of each block $b$ of unit $u$ |